# Consciousness as Intrinsic Structure: Towards a Chemistry of Experience

Matteo Grasso, Jeremiah Hendren, Giulio Tononi

**Abstract:**

To be conscious is to have an experience—not a collection of phenomenal atoms, but a structured whole composed of distinctions and the relations that bind them. Integrated Information Theory (IIT) identifies the essential properties of every experience (axioms), formulates them operationally as postulates that a substrate must satisfy, and unfolds the cause–effect power of the resulting complex into a Φ-structure. Accounting for a content of experience is then a matter of identifying the phenomenal distinctions and relations that compose it, and showing them reflected one-to-one in the causal distinctions and relations of the corresponding Φ-structure. We apply that method to three pervasive contents whose structure is partly open to introspection. Spatial extendedness is composed of spots, whose elemental signature is reflexivity, bound by reflexive inclusion, connection, and fusion; temporal flow is composed of moments, whose signature is directedness, bound by directed inclusion, connection, and fusion; objects bind a particular configuration of features to a general concept through relations bearing the signature of hierarchy. The endeavor is akin to chemistry, which accounts for an endless variety of compounds from a limited set of elements and the ways they bond. We assess these accounts against seven criteria of a good explanation—scope, synthesis, specificity, self-consistency, system consistency, simplicity, and scientific validation—and sketch the prediction that follows: altering the structure specified by a substrate should alter the corresponding content, even when activity and behavior are held comparable. The same method may reach narrow qualia, which resist introspection, and the compound contents that bind many qualia together, though there it remains a proof of concept and an open research program.

## 1. Introduction

To be conscious is to have an experience. It is usually a scene populated by many contents—say, the black of the letters you are reading, the words they form, the white of the page around them, the sounds you are hearing, the feeling of time flowing, sensations from the body, and a background of fleeting thoughts and emotions. Each of these contents can be distinguished from others within the current moment of experience: the black from the white, the letters from the page, the sounds from the silence. Though distinct, these contents are also related: the black letters are in the center of your visual field, they combine into words, each sound lasts a brief moment, and so on. An experience[1] is thus not a collection of phenomenal atoms but a structured whole—what we call a *phenomenal structure*.

This is the starting point of Integrated Information Theory (IIT; see Box: IIT in a box): an experience is a phenomenal structure composed of *phenomenal distinctions* and the *relations* that bind them (**Fig. 1A**). The claim is not merely that experiences can be described structurally from the outside, but rather that an experience itself *is* an intrinsic phenomenal structure; to characterize the quality of experience is therefore to characterize its distinctions and relations. A given experience (or content within it) feels the way it feels because of the particular intrinsic phenomenal structure that makes it what it *is*—in an *absolute* sense, not relative to something else.

[1]Note that in IIT "an experience" always refers to the shortest experiential snapshot or moment. This implies that it should not be confused with a process unfolding in clock time (Comolatti et al., 2025; Tononi & Boly, 2025): a moment of experience may include temporal content—a "feeling of succession"—but it is not itself a "succession of feelings" (James, 1980).

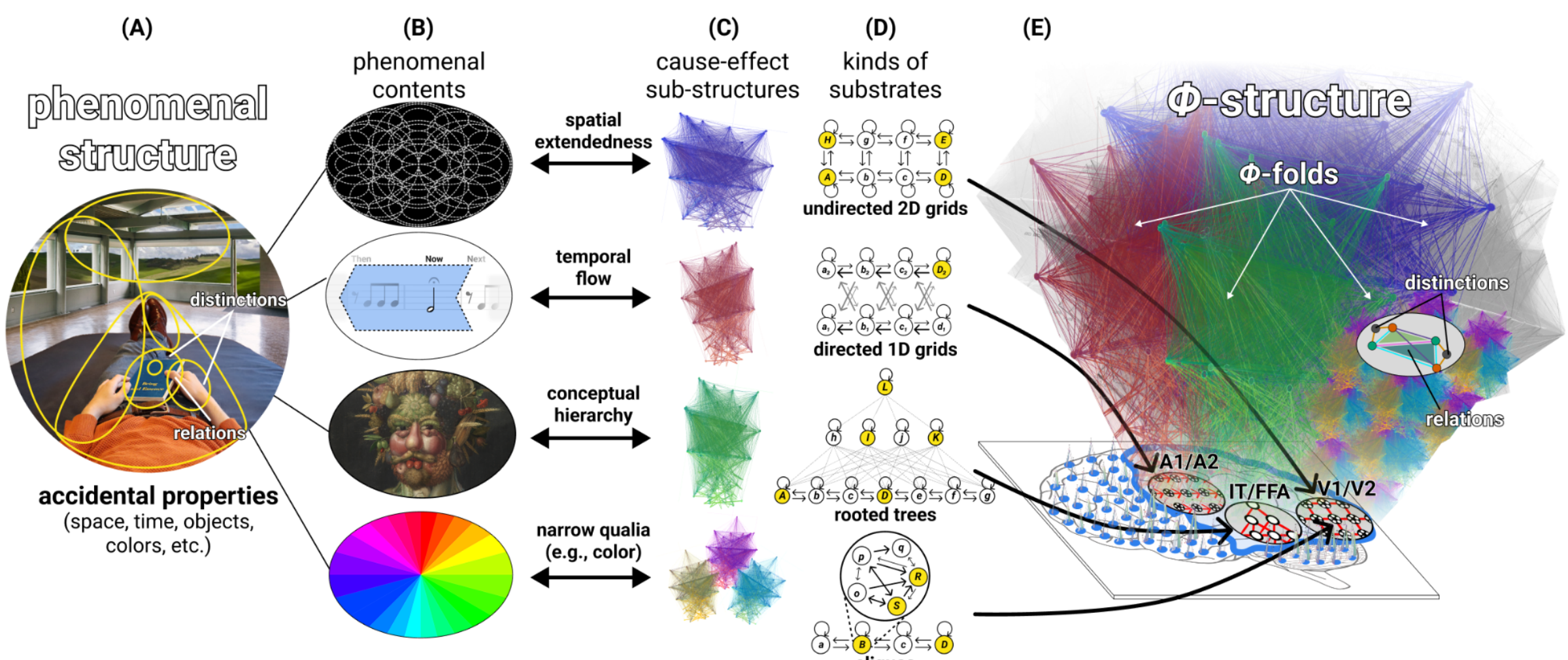

***Figure 1: The methodology and explanatory identity of IIT.*** An experience (A) is an intrinsic phenomenal structure composed of phenomenal distinctions (e.g., book, hand) and the relations that bind them (e.g., hand on book). IIT conjectures an explanatory identity between a phenomenal structure and the corresponding *Φ*-structure unfolded from its substrate (E), which should account for all phenomenal contents with no additional ingredients needed. IIT's account focuses on four kinds of contents (B): spatial extendedness (which characterizes our experience of, e.g., visual and somatosensory space), temporal flow (e.g., the feeling of moments flowing in the extended present), conceptual hierarchy (e.g., the binding of particular configurations of visual features with the general concept that we are seeing "a face"), and narrow qualia (e.g., the hue of specific colors, the timbre of specific sound, and so on). Each kind of content is conjectured to be accounted for by a kind of causal substructure (or *Φ*-fold, C) unfolded from a kind of substrate constituting the main complex (D and E). Undirected 2D grids—reminiscent of the connectivity of early visual cortex (V1/V2) and plausibly somatosensory cortex—should unfold into *Φ*-folds accounting for the defining properties of spatial extensions (Haun & Tononi, 2019); directed 1D grids—reminiscent of the connectivity of early auditory cortex (A1/A2) and plausibly other areas contributing temporal contents (such as MT/V4 for visual motion)—should unfold into *Φ*-folds accounting for the defining properties of temporal flow (Comolatti et al., 2025); rooted trees—reminiscent of the connectivity patterns in the ventral stream of visual cortex (and culminating in IT/FFA) and likely other areas containing invariant mechanisms—should unfold into Φ-folds accounting for the defining properties of conceptual hierarchy; finally, cliques—small groups of highly interconnected units, reminiscent of the ones found in early visual cortex and likely other primary sensory areas such as somatosensory and auditory cortex—should unfold into Φ-folds accounting for the defining properties of narrow qualia (e.g., color, timbre, smell, etc.).

Yet this unique structure is also the basis for why it differs from other experiences (or contents), each with their unique structures.

The goal of IIT is not only to characterize each experience as an intrinsic phenomenal structure but to provide an account of it in scientific terms—as a *Φ-structure* specified by its substrate (**Fig. 1E**). This is done by formulating the axioms of phenomenal existence—the essential properties of every experience—as postulates of physical existence, where "physical" is understood operationally as cause–effect power (Albantakis et al., 2023; Tononi & Boly, 2025). The first four postulates permit the identification of the substrate of consciousness (the *complex*), and by "unfolding" the cause–effect power of its subsets according to the fifth, composition, one obtains its *Φ*-structure. Each causal distinction is the irreducible cause and effect specified by a subset of units, and each causal relation is an overlap among the causes and effects that distinctions specify.

IIT conjectures an *explanatory identity* between a phenomenal structure and the corresponding *Φ*-structure (IIT Wiki/Identity): In causal terms, the quantity of experience is captured by its *Φ* value, while its quality—contents of experience such as spatial extendedness, temporal flow, objects, colors, sounds, thoughts, emotions, and so on (IIT Wiki/Contents)—is captured by the countless

"shapes" a $\Phi$-structure can assume. From the essential properties of experience, formulated in causal terms, everything else follows, with no additional ingredients.

In accounting for the contents of experience, IIT starts by identifying their building blocks and characterizing how they bind. From a few kinds of distinctions and relations, an immense variety of contents can be composed. In this, the endeavor can be seen as akin to chemistry, which accounts for an endless variety of compounds by identifying a limited set of building blocks (the elements) and the ways they combine through relations (chemical bonds). Accounting for a content thus becomes a matter of first identifying the phenomenal distinctions and relations that compose it, and then showing these reflected one-to-one in causal distinctions and relations within the corresponding $\Phi$-structure. The analogy with chemistry, here and in what follows, is useful albeit imperfect: like many analogies, it becomes misleading if taken too far.

The first aim of this chapter is to give a high-level demonstration of how the IIT method applies to three kinds of contents—spatial extendedness, temporal flow, and the conceptual hierarchy that characterizes objects (**Fig. 1B**).[2] The second aim is to generalize from these individual accounts to highlight the foundational principles of a "chemistry of experience": to show how IIT's notion of *intrinsic structure* opens the door to a self- and system-consistent way to both characterize contents of experience and account for them in objective terms. Note that the explanations that follow are brief, and they presume some familiarity with IIT. Complete, multimedia explanations can be found on the IIT Wiki (IIT Wiki/Contents), along with a glossary IIT terms (IIT Wiki/Glossary).

## 2. Where do we start

To build a chemistry of experience, we should not begin with the most intricate compounds but with simple ones that are common and everywhere—the equivalent to carbon dioxide, say, rather than a protein. And we proceed in steps. First, we identify the elements composing that compound: carbon and oxygen, in our analogy. Then we ask what makes each element what it is—its "elemental signature": carbon with four electrons in its outer shell; oxygen with six. This difference in structure is what lets each element arrange, on its own, into different forms: carbon into the flat sheets of graphite or the tetrahedral lattice of a diamond, oxygen into two-atom dioxygen or three-atom ozone. It is what lets the two bind into a further, multi-element compound: one carbon double-bonded to two oxygens, carbon dioxide.

We can ask the same of experience. What are simple, pervasive contents that we can decompose through introspection?[3] IIT proposes to start from spatial extendedness, temporal flow, and objects. We are used to treating these contents as the neutral stage on which other qualia appear; but the extendedness of space, the flow of time, and the invariance of objects are qualia in themselves—just as qualitative and in need of explanation as, say, color or pain. When we see colors arranged in space, their spatial arrangement is not a property of color per se, but acquired by being bound to a spatial structure. If that spatial structure were not there, colors could not feel laid out in an extended field but all superimposed to one another, more akin to odors. Space is thus as much a quality in need of explanation as the colors it locates. Similarly, a melody is a string of notes bound to various moments in time; without temporal flow, there would be no melody, only the experience of sound in the ever-vanishing now. And there is a clear phenomenal difference between seeing unfamiliar ink marks and seeing them forming letters composing a word—the difference made by experiencing them as forming an invariant object.

[2] The account of spatial extendedness has been presented in greater detail in Haun & Tononi (2019, 2025), the one of temporal flow in Comolatti et al. (2025), while the account of conceptual hierarchy is still under development. For further explanations and tutorials also see IIT Wiki/Contents.

[3] Historically, Titchener attempted an inventory of elementary sensations, reaching a total of around 44,000: 32,820 visual, 11,600 auditory, 4 gustatory, plus a small residue of cutaneous, organic and kinesthetic (Titchener, 1896, building on Külpe's *Grundriss der Psychologie*, 1893, which he had translated). Yet the kinds were few: two elements—sensation and affection—varying along five attributes, from which the whole variety was to be generated by combination.

These three contents are an especially good place to start because—unlike color, pitch, or pain—their structure is not only rich but also partially open to introspection (see **Ch. 11, Box: How does IIT use introspection?**). We can attend to a region of space and resolve its parts, dissect a moment into the shorter periods that compose it, or notice how marks form into a familiar object. They are, in this sense, ideal candidates for IIT's method: clearly structured, dissectible from within, and basic enough to be decomposed into elemental components—much as a chemical compound can be decomposed into its elements.

And as with the chemical elements, we can then ask what makes each of these contents different—what building blocks compose them—and how they combine into something further. As we will see, the elemental components of space are *spots*, their elemental signature is *reflexivity*, and when many spots combine through *reflexive inclusion*, *connection*, and *fusion*, they form a *spatial extension*. Time is instead a different phenomenal compound, its elemental components are *moments*, whose elemental signature is *directedness*, and when many moments combine through *directed inclusion*, *connection*, and *fusion*, they form a *temporal flow*. Objects are different still; they are made of *configurations* and *concepts*, characterized by *hierarchy*, and when they combine through *hierarchical inclusion*, *connection*, and *fusion*, they form a *conceptual hierarchy*. And finally, much as carbon and oxygen bind into carbon dioxide, all these phenomenal components can bind into richer compounds—for instance, the experience of seeing a word, in its spatial location, and experiencing it at the end of a sentence.

If we can demonstrate IIT's explanatory identity for these pervasive and introspectable contents, and provide empirical validation, this IIT's characterization of consciousness as intrinsic structure, and the prospect of developing a chemistry of experience. On this basis, we can then develop additional ways to account for contents that are less amenable to introspection (e.g., narrow qualia), for instance, by relying more heavily on inferences in the opposite direction—from properties of the $\Phi$-structure to properties of the phenomenal structure.[4]

### 3. Space, time, and objects in a nutshell

For space, time, and objects, the strategy is the same in each case. First, we begin with a phenomenological analysis: What are the basic elements that compose this content? What relations bind them? What makes each feel the way it does (**Fig. 1B**)? Second, we formulate these properties in causal terms. The question becomes, What kind of causal distinctions and relations can mirror the phenomenal ones (**Fig. 1C**)? And what kind of substrates—and in what state—can specify them (**Fig. 1D**)? (For a more in-depth, multimedia presentation of these accounts, see IIT Wiki/Contents.)

[4] See **chapters 10 and 11** for a discussion of how the Qualia Structure paradigm can contribute to this investigation.

Consider reading the sentence: "All quality is structure." The words lie side by side across the page, each with its location; you take them in one after another, landing on the last while those that preceded it linger in your experience; each word is made of letters, and each letter of smaller marks; and every mark is black against the white page. How do we begin to analyze this experience?

### *3.1* ***Spatial extendedness***

When you see the word "structure" on the page, there is a spatial quality to it (**Fig. 2**, bottom-left). What makes it feel extended—occupying a region, at that location, at that distance from all other regions of space? The word must be embedded in a *visual space* made of components we can call *spots*: phenomenal distinctions composing regions of space, large or small, here or there. A spot may be as large as the whole field or as small as the smallest region one can distinguish. Seeing words does not create space; it makes salient spots that were already there.

The elemental signature of a spot is *reflexivity* (**Fig. 2**, top-left): each spot overlaps or "points to" itself. By this we mean that phenomenal space is static, a persistent sense of what is "here" and "there"—in contrast to, say, the feeling of time, where moments "point away" from themselves. In causal terms, a spot is a causal distinction specified by a subset of units whose receptive fields cover that region, and reflexivity is its causal *motif*: the distinction specifies its cause and its effect over the *same units* (**Fig. 2**, top-right).

But a spot cannot feel extended in isolation; it must relate to other spots in specific ways (**Fig. 2**, top-left). If it did not *include* other spots, it would feel like a point rather than an extended region; if it were not *included* by larger spots, it would feel like the whole field rather than a region within it. If neighboring spots did not *connect*—overlap on a shared region that is itself a spot—they would feel unrelated, at no definite distance from one another. And if connected spots did not *fuse* into a further spot including both and nothing else, space would feel fragmented rather than continuous. Inclusion, connection, and fusion are thus the relations that bind spots into a single, continuous extended field.

In causal terms, each of these relations must hold among distinctions (**Fig. 2**, top-right): a distinction *includes* another when its cause and effect purviews include all the purview elements of the other; two distinctions *connect* when their purviews partially overlap and the overlap is a distinction; and they *fuse* when yet another distinction includes both and nothing more. Because spatial distinctions are reflexive—their cause and effect is over the *same units*—each relation holds symmetrically across cause and effect: these are *reflexive* inclusion, connection, and fusion bonds, unlike the *directed* bonds that will characterize time.

Just as carbon's four valence electrons are its defining elemental signature, allowing it to bind to other atoms in specific ways, a spot's reflexivity is its defining elemental signature, which makes it bind to other spots through reflexive inclusion, connection, and fusion to compose a *spatial extension* (**Fig. 2**, bottom-right). And just as carbon atoms yield graphite or diamond depending on how they are arranged, spots can yield different kinds of space: a one-dimensional space can be unfolded from a one-dimensional grid, as in our example (**Fig. 2**, right), or the two-dimensional space of the visual field can be unfolded from hexagonal grids resembling visual cortex (Haun & Tononi, 2019).

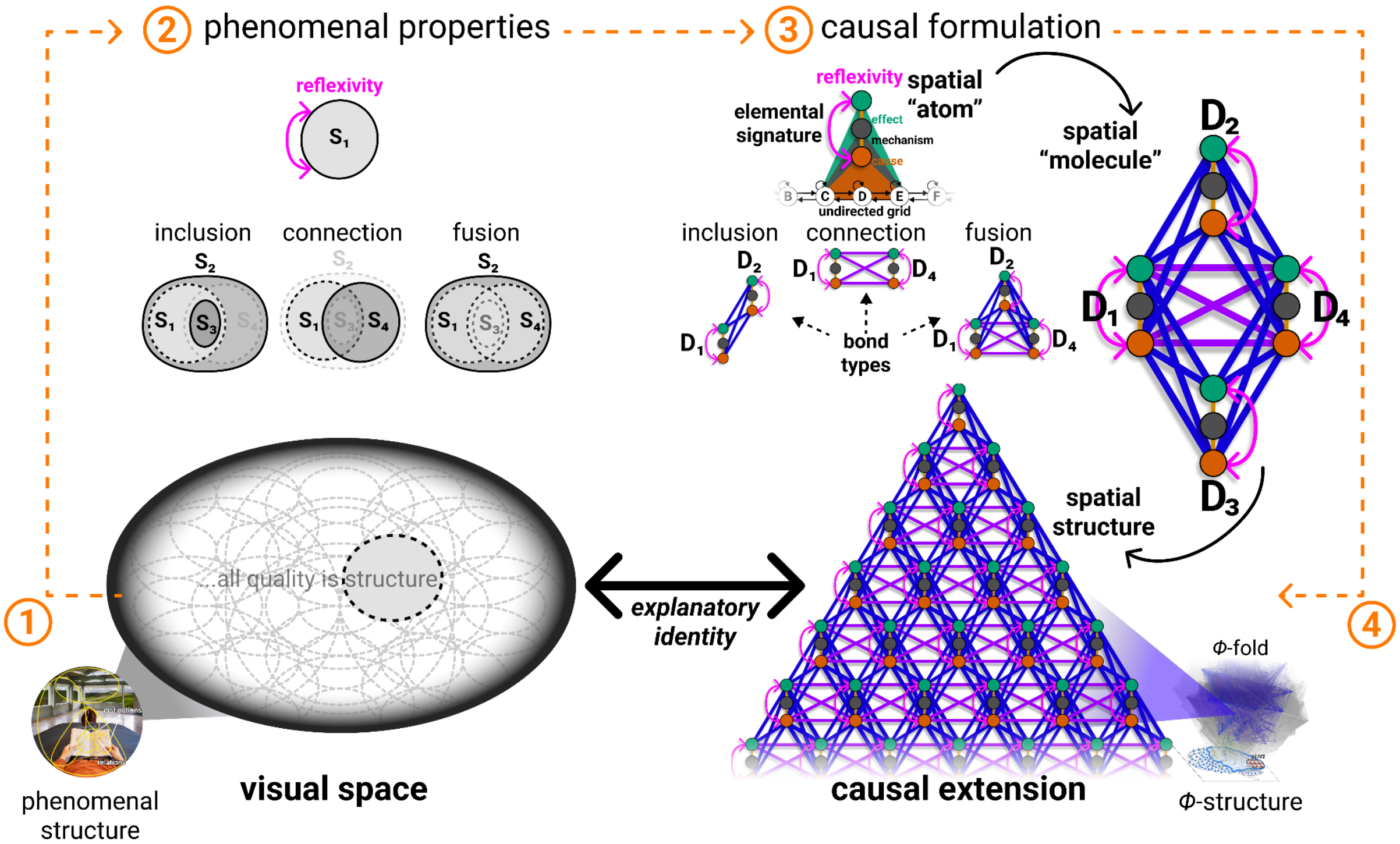

*Figure 2: IIT's account of spatial extendedness.* **(1)** Following the method of IIT, we identify a phenomenal content—here it's the extendedness of visual space—abstracted away from all other contents (e.g., the feeling of our body). We then focus on an elementary component of this content, for instance a spot (here, the one occupied by the word "structure"). **(2)** We characterize the basic phenomenal properties and relations that make this component feel extended: A spot ($S_1$) is reflexive (point to itself); it is included and it includes other spots ($S_2$ and $S_3$); it connects to other spots ($S_4$); and it fuses into other spots ($S_2$), all in a reflexive way. **(3)** We formulate these properties and relations in causal terms. The causal distinctions unfolded from an undirected grid are the atoms of a spatial structure because they are reflexive: they specify a cause and effect over the same units. In our example, mechanism CDE (indicated in black) specifies both a cause (in red) and an effect (in green) over the same units, CDE (fuchsia relation for "full overlap"). Reflexive distinctions bond through reflexive inclusion when the purviews of one distinction (here $D_1$) are fully included by the purviews of the other ($D_2$) (indigo for "inclusion" bonds); through reflexive connection when their purviews partially overlap ($D_1$ and $D_4$) (purple for "partial overlap" bonds); and through reflexive fusion when a third distinction ($D_2$) includes both ($D_1$ and $D_4$) and nothing else (through inclusion and partial overlap bonds). Small substructures composed of distinctions that satisfy all four properties (e.g., $D_1$ and $D_4$) start to feel extended (they are like spatial "molecules"). **(4)** When many reflexive distinctions bind through reflexive bonds, they compose into a spatial structure called a causal extension—a *Φ*-fold within the full *Φ*-structure—corresponding to a content within the full phenomenal structure. The success of this account provides validation to the explanatory identity between phenomenal structure and Φ-structure.

The spatial quality of seeing a word on the page is hence given by the spot it occupies and its relations to the rest of the space.

### *3.2 **Temporal flow***

When you read the word "structure" at the end of the sentence, there is also a temporal quality to it (**Fig. 3**, bottom-left). What makes it feel flowing in time—lasting a certain period, at that temporal location, occurring now and succeeding the words that just passed? The word must be embedded in an *extended present* made of components we can call *moments*: phenomenal distinctions composing periods of time, short or long. A moment may be as long as the whole present or as short as the smallest period one can distinguish. The words do not create time; they make salient moments that were already there.

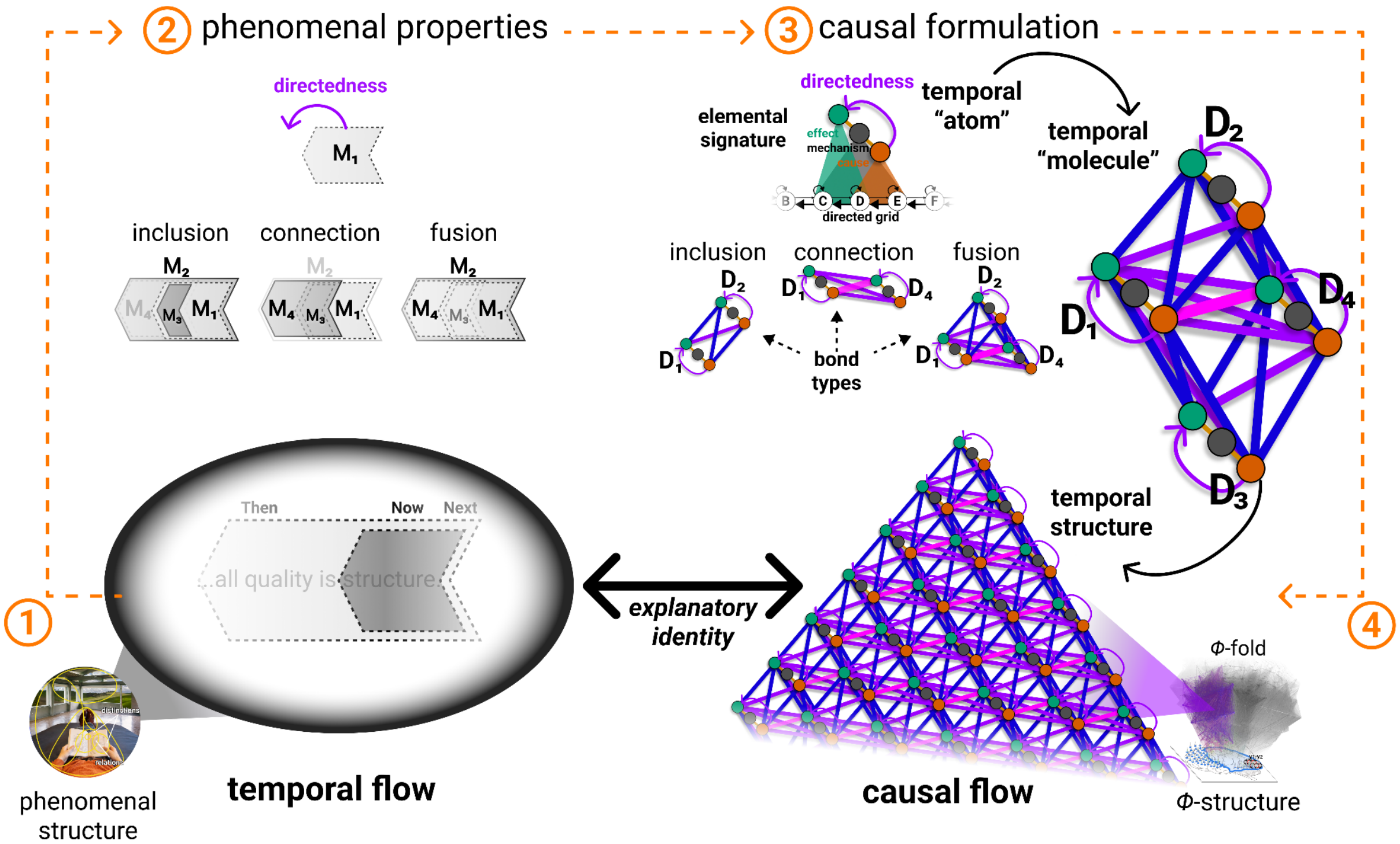

***Figure 3: IIT's account of temporal flow.*** **(1)** We identify a phenomenal content—temporal flow—abstracted away from all other contents. We then focus on an elementary component of this content, for instance a moment (here, the one occupied by reading the word "structure"). **(2)** We characterize the basic phenomenal properties and relations that make this component feel temporal: A moment (M1) is directed (points away from itself); it is included and it includes other moment (M2 and M3); it connects to other moments (M4); and it fuses into other moments (M2), all in a directed way. **(3)** We formulate these properties and relations in causal terms. The causal distinctions unfolded from a directed grid are the atoms of a temporal structure because they are directed: they specify a cause and effect over different units—cause towards the "now" and effect towards the "then" terminus of the grid. In our example, mechanism CDE (indicated in black) specifies a cause (in red) over DE and an effect (in green) over CD (this time purple, for partial overlap). Directed distinctions bond through inclusion when the purviews of one distinction (here M1) are included by the purviews of the other (M2), this time directed—with three inclusions (indigo) and one partial overlap (purple) bond; through directed connection when their purviews partially overlap (M1 and M4)—but with an extra full overlap bond; and through directed fusion when a third distinction (M2) includes both (M1 and M4) and nothing else—with inclusion, full and partial overlap bonds, you get it by now. Small substructures composed of distinctions that satisfy all four properties (e.g., M1 and M4) start to feel flowing (they are like time "molecules"). **(4)** When many directed distinctions bind through directed bonds, they compose into a temporal structure called a causal flow—a Φ-fold within the full Φ-structure—corresponding to the temporal flow within the full phenomenal structure. The success of this account provides further validation to the explanatory identity between phenomenal structure and Φ-structure.

The elemental signature of a moment is *directedness* (**Fig. 3**, top-left): each moment feels to "point away" from itself, from the *now* toward the *then*, slipping toward what has just passed. In causal terms, a moment is a causal distinction specified by a subset of units covering that period, and directedness is its causal *motif*: the distinction specifies its cause and its effect over *different* units—the cause toward the *now*, the effect toward the *then* (**Fig. 3**, top-right).

A moment, too, cannot feel extended in isolation; it must relate to other moments (**Fig. 3**, top-left). If it did not *include* other moments, it would feel like an instant rather than lasting a certain period of time; if it were not *included* by larger moments, it would feel like the whole present rather than a moment within it. If successive moments did not *connect*—overlap on a shared period that is itself a moment—they would feel unrelated, at no definite interval from one another. And if connected moments did not *fuse* into yet another moment including both and nothing else, time would feel fragmented rather than continuous. Inclusion, connection, and fusion are the relations that bind moments into a single, continuous flow. In causal terms, these relations hold among distinctions exactly as for space—with one difference: because temporal distinctions are *directed*, with cause and effect over different units, each relation holds in a directed way. These are *directed* inclusion, connection, and fusion bonds. Directed connection, in particular, orders the moments: the effect of one overlaps the cause of the other, so that one is felt as the *predecessor* and the other as the *successor* (**Fig. 3**, top-right).

Just as oxygen's six valence electrons let it bind to other atoms in specific ways, a moment's directedness makes it bind to other moments through directed inclusion, connection, and fusion to compose a *temporal flow* (**Fig. 3**, bottom-center). And just as oxygen atoms yield dioxygen or ozone depending on how they are arranged, moments yield flows of longer or shorter duration, unfolded from larger or smaller directed grids (Comolatti et al., 2025).

The temporal quality of reading the last word of a sentence is hence given by the moment it occupies and its relations to the rest of the temporal flow.

### *3.3 Objects*

With objects, IIT's method reaches toward a more complex and less introspectable compound. In the visual domain, objects bind spots, moments, and narrow qualia into recognizable wholes—a face, a letter, a word: the black of a letter is not a free-floating hue but the color of *that* letter, in that location, with that shape. This further compounding is often phenomenologically hidden, but it surfaces when we learn a new concept. Before learning to read, we experience mere marks—features bound to regions of space. After learning a single character, we unitize those features into a particular *configuration*, such as "s"; and once we learn its variants (s, *s*, S), we experience each configuration as an instance of a general *concept*—as a token of a type ("*this* s is *an* S"). Once acquired, the mark could not feel like "an S" were it not bound to this hierarchy of alternative S's, much as a spot could not feel "here" without a field of "theres."

IIT thus characterizes an object as a *hierarchy* binding a *particular configuration* of features with a *general concept*. Just as reflexivity is the elemental signature of spots and directedness of moments, *hierarchy* is the elemental signature of objects. But unlike a spot or a moment, this component comes in two types: *configurations*, which bind features into a particular token, and *concepts*, which bind that token to an equivalence class of alternatives, a general type. For example, several curved edges combine into this particular "s" configuration, while our concept "S" binds it to an equivalence class of variants (s, *s*, S). A phenomenal object is the binding of the two—a configuration (the particular) to a concept (the general).

If IIT is correct, objects too correspond to $\Phi$-folds called *conceptual hierarchies*, unfolded from rooted-tree substrates such as the ventral visual stream, where face- and concept-selective cells suggest a hierarchy from features to invariants. Conjunction-like units would specify the distinctions corresponding to configurations, disjunction-like units those corresponding to concepts, and relations of *hierarchical inclusion*, *connection*, and *fusion* would bind features, configurations, and concepts into an object: these contours and colors making up these marks, in turn making up "this s," which feels like "an S".

Configurations and concepts—both bearing the signature of hierarchy—bind through hierarchical inclusion, connection, and fusion to compose a conceptual hierarchy. This account is still under development (Tononi & Boly, 2025; Grasso & Tononi, in preparation), but it shows the same method reaching toward progressively more complex and less introspectable contents of experience.

### 4. A good explanation?

We have now seen the same method applied three times. In each case, we identify the elemental distinctions that compose a content, their elemental signature—reflexivity, directedness, or hierarchy—and the bonds through which they combine—reflexive, directed, or hierarchical inclusion, connection, and fusion. These are only initial characterizations of what the full accounts will look like. But let us imagine them complete, so that space, time, and objects can be understood as different $\Phi$-folds unfolded from different substrates. Does this amount to a good explanation of these contents of experience?

Drawing inspiration from criteria long proposed in epistemology and the philosophy of science (e.g., Popper, 1959; Kuhn, 1977; Quine & Ullian, 1970), we can summarize what characterizes a good explanation in seven S's: : scope, synthesis, specificity, self-consistency, system consistency, simplicity, and scientific validation (IIT Wiki/Method FAQs). These are much the same considerations by which we judge the explanation provided by chemistry to be satisfactory, and each points to something the others do not.

*Scope*. Chemistry accounts for substances across an enormous range—from simple gases and salts to crystals, polymers, and the compounds of living cells. IIT's proposed chemistry of experience reaches across contents that are just as diverse. Scope applies within the accounts sketched above—for example, the account of space also covers how we experience regions, distances, locations, sizes, and boundaries (Haun & Tononi, 2019); likewise the account of time also covers how we experience periods, durations, intervals, and temporal boundaries (Comolatti et al., 2025). More importantly, however, it applies far beyond the three contents discussed here—from other contents such as color, sound, pain, and emotions (and the compound contents that bind them), to an account of the quantity of consciousness: why the posterior cortex supports it while the cerebellum, with far more neurons, does not; why it fades in dreamless sleep and under anesthesia, though neurons remain active; and, by extrapolation, which systems beyond ourselves may be conscious, whether biological or artificial.

*Synthesis*. Chemistry shows how apparently disparate properties arise from common compositional principles. The hardness of diamond and the softness of graphite, for example, belong to a unitary account once the properties and different arrangements of carbon atoms are specified. IIT likewise brings extendedness, flow, objecthood, and every other experiential content under the explanatory principle of intrinsic structure: phenomenal distinctions and relations are identical to causal distinctions and relations within a $\Phi$-structure. What differs among contents is not the form of explanation but the elemental signature and binding patterns that shape their intrinsic structure.

*Specificity*. Chemical explanations derive particular properties from molecular structure rather than merely noting that a substance has properties. Isomers with the same kinds and numbers of atoms differ because those atoms are arranged differently. IIT's approach is similarly committed to detail.

Reflexivity and reflexive bonds, for example, must account for every nuance of spatial experience—why a spot feels *here*, with *this* size and boundary, connected to neighboring spots at *these* distances. Because the proposed identity is one-to-one, changing a causal distinction or relation should change the corresponding experiential content in specific ways.

*Self-consistency*. Chemical terms retain their meaning from one compound to another: an atom, a valence, or a bond cannot be defined one way for a crystal and another for an organic molecule. In IIT, the same principles govern the analysis of experiential contents—whether of space, time, or objects. At the same time, consistency does not mean treating unlike structures as interchangeable. Reflexivity accounts for the symmetric organization of space, directedness for the asymmetric order of temporal flow, and hierarchy for the binding of particular configurations to general concepts. Each account must therefore fit the common framework without borrowing the structural signature of another content.

*System consistency*. Molecular explanations fit into our greater scientific picture: chemical bonds depend on physical interactions, while molecular organization helps account for the behavior of materials and living systems. IIT's account must likewise cohere with our understanding of the natural world, yet the burden of system consistency is even greater when our explanatory target is consciousness and its contents: we must connect claims about phenomenology to claims about causal and neural organization, without changing explanatory vocabularies midway. This is made possible, again, through the concept of intrinsic structure—recognized through introspection and operationalized in terms of causal power. The causal motifs that we postulate must be translated into neural motifs, which—if and when validated—will not feel to be arbitrary correlates because they are grounded in a structural form of reasoning derived from phenomenology: undirected grids are suited to spatial structure, directed grids to temporal flow, and rooted hierarchies to the binding of configurations and concepts, and so forth for additional qualia (**Fig. 2D**).

*Simplicity*. The simplicity of chemistry lies in the economy of principles by which it explains a wide variety of substances. New compounds do not require a new kind of matter or a separate law of bonding for each case; different elements, bonds, and arrangements suffice. IIT aims at the same economy. Its entire account of consciousness follows from five essential properties of experience, translated into causal terms by the postulates, with no need for additional ingredients. This applies to both the quantity of experience and its contents, where the same postulates apply; what changes is the substrate architecture, its state, and thus the distinctions and relations it specifies.

*Scientific validation*. Molecular structure is explanatory because it supports tests: changing a molecule's composition or bonds yields definite changes in its reactions and macroscopic properties. IIT's chemistry of experience must answer to the same demand. If a content is identical to a $\Phi$-fold specified by the substrate—and not with its activity or function—then IIT predicts that altering that structure should alter the content even when activity and behavior are held comparable; and failing to find this would count against IIT. For space, time, and objects, this prediction takes different forms.[5]

---

[5] The same demand applies to the presence of consciousness, where the account is furthest along. It explains why consciousness depends on the corticothalamic system and not on the cerebellum, whose largely independent, feedforward micro-zones cannot constitute a large complex despite four times as many neurons; why it is lost in dreamless sleep, anesthesia, and some seizures, when bistable or paroxysmal activity breaks down integration though the cortex remains active; and why the cortex can divide into more than one complex, as in the split brain (Tononi et

For space, IIT predicts that co-activating nearby locations of the visual field should distort perceived distance, suggesting that a change in connectivity alone can reshape spatial experience (Song et al., 2017). A cortical scotoma should not merely blank out a region but warp the space around it, as surrounding regions are "sewn together"—a prediction under test in the INTREPID adversarial collaboration (Corcoran et al., 2026). And because inactive units may still take and make a difference, silencing them without rendering them unable to fire should alter perceived extension, whereas truly inactivating them should not (Takahashi et al., 2025); the same reasoning may bear on the pure extendedness reported in deep meditation (Boly et al., 2024).

For time, the substrate of the extended present should correspond to a single macro-state of directed grids rather than to a succession of neural events unfolding in clock time; the duration of the present should scale with the number of units composing the grid, and that of the instant with their update grain; activation toward the now- or then-terminus should shift whether a stimulus is experienced as happening now or then; and changes in connectivity should speed up or slow down the temporal flow independently of activity, as in strong emotion or altered states (Comolatti et al., 2025).

For objects, rooted-tree hierarchies such as the ventral visual stream should support the invariants that conceptual hierarchies require, and account for the contribution to experience of the face- and concept-selective cells found there; and conceptual hierarchies should be selectively disruptable—a lesion of high-level areas impairing the experience of an object as such while sparing its features—as clinical dissociations already suggest (Grasso & Tononi, in preparation).

Taken together, these seven criteria mark the difference between building a catalogue of contents of experience and accounting for them; the same difference between the periodic table—that merely lists the elements—and chemistry—that explains how they combine.[6] How far that chemistry can reach is the question we take up next.

## 5. Where do we go from here

We have demonstrated how to account for contents whose structure is at least partly possible to decompose through introspection. But experience is far richer. Beyond phenomenal space, time, and objects are numerous *narrow qualia*—the greenness of a blade of grass, the pitch of a tone, the painfulness of a prick—and *broader qualia* that bind many contents together (Balduzzi &

al., 2016; Pigorini et al., 2015; Sasai et al., 2016). It predicts that the complexity of the cortical response to direct perturbation should track consciousness rather than behavior—high in dreaming as in wakefulness, low in dreamless sleep and anesthesia—which is what the perturbational complexity index measures, and what stratifies unresponsive patients at the bedside (Casali et al., 2013; Casarotto et al., 2016). And it predicts that posterior cortex should suffice, without prefrontal broadcasting or monitoring, as tested in the ARC-COGITATE adversarial collaboration (Cogitate Consortium, 2025).

[6] A closely related effort is the Qualia Structure paradigm (Tsuchiya, 2025), which explicitly aims to compile a "periodic table of qualia": a systematic tabulation of the kinds of qualia and the relations among them, established third-person by comparing experiences through behavioral similarity judgments—relations that are nonetheless taken to reflect the structure of experience itself. The Qualia Structure paradigm characterizes relations extrinsically, and so far tabulates the kinds of qualia without yet specifying the principles by which they compose into one another—which leaves to future work, to be approached through category and sheaf theory. For IIT instead the elements are specific kinds of causal distinctions and the bonds are causal relations, both intrinsic to the causal power of the substrate, which already furnish the principles by which contents are composed.

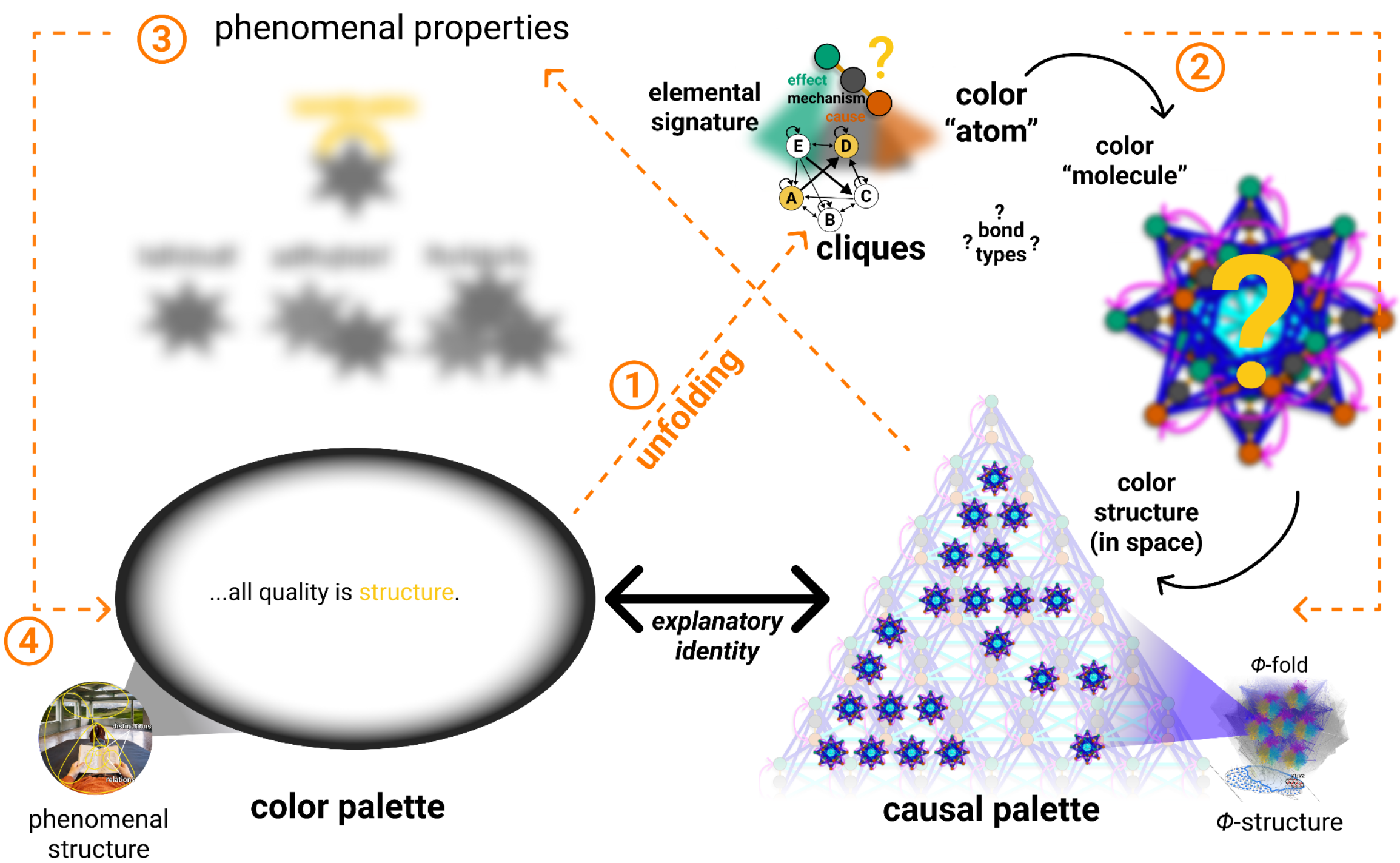

***Figure 4: A speculation on the approach to narrow qualia.*** We again identify a phenomenal content—in this case color—abstracted away from all other contents. Since narrow qualia like color are impenetrable to introspection, we cannot characterize its basic phenomenal properties. **(1)** We thus immediately resort to unfolding its substrate (e.g., *cliques*) into *kernels* (color "atoms") to discover elemental signature and bonding types. **(2)** We would study how these kernels relate to one another to form more complex substructures (like binding to visual space). **(3)** We then utilize this knowledge to gain insight on the internal structure of phenomenal color, and to design experiments to validate this account (or resort to relational methods, e.g., from Qualia Structure, see Ch. 11). **(4)** The success of this account would provide even further validation to the explanatory identity between phenomenal structure and $\Phi$-structure, and it would constitute compelling evidence that all quality is structure.

Tononi, 2009; Kanai & Tsuchiya, 2012; Lee-Youngzie et al., 2026). Can the same method reach them? Here we can only indicate the direction proposed by IIT.

Narrow qualia are the hardest case. While we can tell red from orange, or judge red to be more similar to orange than to blue, we cannot introspectively decompose the redness of red into distinctions and relations. This is why narrow qualia have long been taken as paradigmatic examples of the hard problem (Chalmers, 1995)—their quality seeming unstructured and so ineffable.

To account for narrow qualia, IIT proposes to proceed in the "opposite" direction—from knowledge about the substrate of experience to the experience itself. Having built an account that satisfies the seven S's where introspection can validate it, we can make an *inference "from" a good explanation* (Tononi & Boly, 2025): if the explanatory identity holds for space, time, and objects, and we have reasons to believe that "all quality is structure," then a narrow quale too must correspond to an intrinsic structure, even if it resists introspection. IIT conjectures a type of structure called a *kernel*, unfolded from a tightly interconnected *clique* of units, such as the color-opponent minicolumns repeated across the visual cortex (Tononi & Boly, 2025).

Introspection cannot reach this kernel: its elemental signature and the bonds that compose it lie beneath what we can attend to. But we can study the clique of units that specify the kernel, and the recovered structure may in turn illuminate the phenomenology—for instance, by suggesting interventions that should alter the quale (**Fig. 4**). Other methods can help as well: the Qualia Structure paradigm could complement IIT's approach by probing a narrow quale through its relations of similarity and difference to other qualia, much as chemists characterize an element through how it reacts, or the light it emits, rather than by inspecting the atom directly (see **Ch. 11**).

So far we have engaged with something akin to molecular chemistry, only focusing on basic components like atoms and molecules. But most phenomenal contents are compounds far more complex—less like carbon dioxide and more like proteins. Such *compound contents* include visual motion; the multimodal space of the body and the world around us; and perhaps emotions, thoughts, and the sense of self (Tononi & Boly, 2025). The feeling of a bee sting, for instance, may bind a burning tactile quality, a location in bodily space, a duration in the present, a negative valence, and an urge to withdraw—a compound of kernels, extensions, flows, and hierarchies. If our chemistry accounts for the basics and proves a good explanation there, we can then envision ways it might tackle contents of this complexity too. For now, IIT's account of narrow and compound qualia is best regarded as a proof of concept and an open research program rather than a complete explanation: unfolding systems is still computationally prohibitive, and the mapping from structure to phenomenology must ultimately be validated against a phenomenology that is itself hard to access. Even so, it is a proof of concept of a kind never before attempted, let alone achieved—an account of qualia that specifies what would count as an explanation in the first place: not a neural correlate, an activity pattern, or a function, but an intrinsic causal structure that corresponds in every detail to the intrinsic phenomenal structure that defines the way the experience feels.[7]

[7] See **Ch. 11** for a discussion of what it means to characterize experience intrinsically rather than extrinsically, and of whether extrinsic methods such as those of the Qualia Structure paradigm can bolster IIT's account. See **Ch. 10** for a discussion on how IIT's account can ground inferences on the experience of other beings.

**IIT in a box**

IIT aims to account for experience in objective, physical terms by answering two questions: why is experience present rather than absent, and why do specific experiences feel the way they do?

A full overview of IIT can be found in the IIT Wiki, together with a list of IIT publications, a glossary of terms, and a list of FAQs.

In summary, IIT's method is "consciousness first" (IIT Wiki/Overview). It starts from the immediate and irrefutable fact that experience exists (IIT Wiki/Foundations), and uses introspection to identify the essential properties of every conceivable experience. These are expressed by the theory's axioms, which characterize experience as intrinsic, specific, unitary, definite, and structured. IIT then formulates these phenomenal properties in physical, operational terms as postulates about the substrate of consciousness (IIT Wiki/Axioms and Postulates).

Applying the postulates to a substrate permits us to identify a *complex*: a set of units in its current state whose cause–effect power upon itself is maximally irreducible. The complex is the substrate of consciousness. Once the complex is identified, its cause–effect power is unfolded into a $\Phi$-structure composed of causal distinctions and relations (IIT Wiki/Unfolding).

The central claim of IIT is the *explanatory identity* between an experience and the $\Phi$-structure specified by its substrate (IIT Wiki/Identity). The quantity of consciousness corresponds to the amount of integrated information, while the quality of consciousness corresponds to the way the $\Phi$-structure is composed. In this sense, IIT argues that *all quality is structure*. Essential properties of experience are accounted for by the postulates themselves; accidental properties—space, time, objects, and narrow qualia—correspond to sub-structures, or $\Phi$-folds, within the overall $\Phi$-structure. This framework yields a research program for contents of experience (IIT Wiki/Contents), beginning with spatial extendedness, temporal flow, and objects.

Empirical validation begins in humans who can introspect and report, and then tests whether the anatomical and physiological properties of brain substrates support the predicted complexes and $\Phi$-structures. IIT has been used to account for the dependence of consciousness on certain parts of the brain and not others, its fading in dreamless sleep and anesthesia, and its return in waking and dreaming (IIT Wiki/Validation). If validated, IIT permits principled extrapolations to unresponsive patients, infants, non-human animals, and artifacts—based on substrate properties rather than intelligence, behavior, or cognitive performance. It also leads to an intrinsic ontology of consciousness and its place in nature, with implications for meaning, values, purpose, and freedom (IIT Wiki/Intrinsic Ontology).